\documentstyle[12pt,epsf,a4]{article}
\newcommand{\bramket}[3]{\left\langle\,{#1}\,\left|\,{#2}\,
            \right|\,{#3}\,\right\rangle}
\newcommand{\bqn}{\begin{eqnarray}}
\newcommand{\eqn}{\end{eqnarray}}
\newcommand{\beq}{\begin{equation}}
\newcommand{\eeq}{\end{equation}}
\newcommand{\CL}{{\cal L}}
\newcommand{\mbf}[1]{\mbox{\boldmath$#1$}}
\newcommand{\ovl}[1]{\overline{#1}}
\newcommand{\bm}[1]{\bibitem{#1}}
\newcommand{\rw}{\rightarrow}

\begin{document}
\renewcommand{\thefootnote}{\fnsymbol{footnote}}
\setcounter{footnote}{1}

\title{\bf Weak Decays of Heavy Mesons in the
  Instantaneous Bethe Salpeter Approach}

\vspace{2cm}
\author{{\large G. Z\"oller, S. Hainzl, C.R. M\"unz}\\[2ex]
{\em Institut f\"ur Theoretische Kernphysik der Universit\"at Bonn,}\\
{\em Nu{\ss}allee 14-16, 53115 Bonn, FRG}\\[2ex]
{\large  M. Beyer}\\[2ex]
{\em Institut f\"ur Theoretische Kernphysik der Universit\"at Bonn,}\\
{\em Nu{\ss}allee 14-16, 53115 Bonn, FRG}\\
and\\
{\em Max Planck Gesellschaft, AG
`Theoretische Vielteilchenphysik'}\\
{\em  Universit\"at Rostock, Universit\"atsplatz 1,
18055 Rostock, FRG}\\[11cm]
{\sf FAX 0381-4982857, Email:beyer@darss.mpg.uni-rostock.de}\\
{\sf BONN TK-94-15, MPG-VT-UR-44/94}\\[-11cm]}
\date{}
\maketitle
\thispagestyle{empty}

{\bf Abstract:} In the framework of the instantaneous Bethe Salpeter
equation we investigate weak decays of $B$ and $D$ mesons. Mesons are
described as $q\ovl q$ states interacting via a mixture of a scalar
and a vector confining kernel and a one gluon exchange. The model
parameters are fixed by a fit to the meson mass spectrum including
also the light mesons.  We calculate form factors and compare our
results to the pole dominance hypothesis. From a fit to ARGUS and CLEO
data on $B\rw D^*\ell\nu$ semileptonic decay we extract the Cabbibo
Kobayashi Maskawa matrix element to be $V_{cb}=(0.032\pm
0.003)(1.49ps/\tau_B)^{1/2}$. The Isgur Wise function is calculated
utilizing the heavy quark mass limit. Finally, we give some results on
non-leptonic decays.

\newpage

\section{Introduction}

In recent years the decay of $B$ and $D$ mesons has become one of the
most exciting source of information on fundamental interactions and
symmetries. More experiments are expected, since the plans for future
$B$-meson factories are now beginning to be realized.

Theoretical effort to extract the fundamental quantities relevant on
the `quark level' from the `hadronic counting rates' has been enormous
and also successful. The ultimate aim is to extract those quantities
model independent. On the other hand, interpretation in terms of model
degrees of freedom has been proven very useful: Since QCD and
confinement is still not solved at moderate energies, weak decays may
be used to improve on the dynamical description of the underlying
quark structure and therefore to extract information on the relevant
degrees of freedom.

Considerable amount of effort has been put into the description of
mesons in terms of the underlying quark structure. For heavy mesons,
e.g. charmonium and bottomonium, non-relativistic models have been
particularly successful~\cite{ruj75}-\cite{lic90}.  For a review see
e.g.~\cite{lic87, luc91}. However, a closer look reveals that
relativistic effects are non-negligible even for mesons involving only
heavy quarks. We found that those are particularly important for $M1$
transitions in charmonium and even in the form factors of $B$-meson
decays~\cite{bey92,res94}. We also found that non-covariance of the
non-relativistic model may lead to inconsistencies for the $B$ decay
form factors of about $5-10\%$~\cite{res94}. The non-relativistic model
is worse in cases where heavy-to-light or light-to-light transitions
are involved. Although the mass spectrum may be reproduced, the form
factors and therefore the description of the decay rates fails
badly~\cite{res93}.

We have now chosen a quite different approach utilizing the
instantaneous Bethe Salpeter equation (IBS) to treat the $q\ovl q$
system within a relativistically covariant formalism
\cite{jres94}. The model is able to describe the meson mass spectrum
for low radial excitations. It has been applied to the calculation of
leptonic decays, viz.  decay constants, $\gamma\gamma$
decays~\cite{mue94}, and to elastic form factors of mesons~\cite{MR94}
as well as to charmonium and bottomonium~\cite{RM94}. Similar lines
have been followed by~\cite{lag92}-~\cite{fau92}. In the following
section we will give a short survey of the model.

To calculate semileptonic decays, we first determine the two form
factors of the $0^- \rw 0^-$ transitions and the four form
factors of the $0^- \rw 1^-$ transitions.  We compare our
results to the pole dominance ansatz, which is a different way to
describe the $q^2$ behavior of the form factors.  This will be
presented in section~3.

Recently, much attention has been paid to heavy quark effective theory
(HQET) (see e.g.~\cite{neu93} and references therein), which relates
form factors of $B\rw D $ to those of $B\rw D^*$
transitions introducing heavy quark symmetries. All form factors are
then related to one universal function (i.e. the Isgur Wise
function~\cite{isg90}). Some of the more recent calculations utilizing
the Bethe Salpeter approach are specifically dedicated to calculate
this universal function~\cite{wam94, kug93}.  In section 4 we connect
our results to the notion of HQET and give the result for the Isgur
Wise function.

In section 5 we discuss light meson transitions. We compare the
Bethe-Salpeter approach to the non-relativistic ansatz. However, since
we use an instantaneous kernel and ladder approximation, results for
the light mesons should be less reliable as for the heavy mesons.

Section 6 is dedicated to non-leptonic decays of heavy mesons.  The
discussion on QCD corrections in the operators, initiated by the
latest CLEO results on $B$ decays is not entered here. Through being
an important issue it is not considered the main subject of the
present paper. For definiteness we neglect additional gluonic
corrections in the weak operators.

We conclude with a summary of our main results in section~7.

\section{The quark model}
Mesons are treated as $q\ovl q$ states in the framework of the
instantaneous Bethe Salpeter (BS) equation. A more detailed
description of the model used here, has been given in~\cite{jres94,
mue94} in a different context, and references therein. Here we
summarize the main results relevant for weak decays of mesons.

For an instantaneous Bethe Salpeter kernel and utilizing free quark
propagators with effective quark masses $m_1$ and $m_2$ one may
perform the $p^0$ integrals of the BS-equation. This is done in
the rest frame of the bound state with mass \(M\) and lead to the
(full) Salpeter equation, viz.
\begin{eqnarray}
\Phi(\mbf{p}) &=&
\int \!\!\frac{d^3p'}{(2\pi)^3}\,
\frac{\Lambda^-_1(\mbf{p})\,\gamma^0\,
[V(\mbf{p},\mbf{p}\,')\,\Phi(\mbf{p}\,')]
\,\gamma^0\,\Lambda^+_2(-\mbf{p})}
{M+\omega_1+\omega_2}
 \nonumber \\
 &-&
\int \!\!\frac{d^3p'}{(2\pi)^3}\,
\frac{\Lambda^+_1(\mbf{p})\,\gamma^0\,
[V(\mbf{p},\mbf{p}\,')\,\Phi(\mbf{p}\,')]
\,\gamma^0\,\Lambda^-_2(-\mbf{p})}
{M-\omega_1-\omega_2}
 \label{salpeter}
\end{eqnarray}
with
\beq
\Phi(\mbf{p})=\int \!\!\frac{dp^0}{(2\pi)}\,
\chi_P(p^0,\mbf{p})|_{P=(M,\mbf{0})}
\eeq
where $\chi_P(p^0,\mbf{p})$ is the full Bethe-Salpeter amplitude.
Here \(\omega_i=\sqrt{\mbf{p}\,^2+m_i^2}\), and we introduce energy projection
operators \( \Lambda^{\pm}_i(\mbf{p}) = (\omega_i \pm
H_i(\mbf{p}))/(2\omega_i) \) in obvious notation,  where
\(H_i(\mbf{p})=\gamma^0(\mbf{\gamma}\cdot \mbf{p}+m_i)\) is the standard
Dirac Hamiltonian (see e.g. refs.\cite{jres94,mue94}).

The dynamical input of the model is defined by a confinement plus one
gluon exchange (OGE)  kernel, $V=V_C+V_G$.
Confinement is introduced as a mixture of a scalar and a
vector  type kernel, viz.
\begin{equation}
\left[V_C(\mbf{p},\mbf{p}\,')\,\Phi(\mbf{p}\,')\right]
= \;\;\;{\cal V}^S_C((\mbf{p}-\mbf{p}\,')^2)\,[\Phi(\mbf{p}\,')
 - \,\gamma^0\,\Phi(\mbf{p}\,')\,\gamma^0]
\label{cc}
\end{equation}
with  a scalar function  \({\cal V}_C\).
The mixture of a scalar and a vector spin structure has been
introduced in order to give an improved description of the spin orbit
splitting. Note, that a purely scalar confining kernel leads to an RPA
instability of the instantaneous Bethe Salpeter equation.

Since is it more suggestive to introduce a confining potential in
co-ordinate space, we assume that the Fourier transform is given by
\beq
{\cal V}_C^F(r) = a_c+b_c r
\label{vconf}
\eeq

To derive the OGE kernel, we chose Coloumb gauge for the gluon
propagator. This way it is possible to retain a covariant formulation
within an instantaneous treatment of the BS equation, and it allows
to substitute \(q^2\) by \(-\mbf{q}^{\,2}\).
 The OGE kernel then reads \cite{tjo90,mur83}
\begin{eqnarray}
\lefteqn{\left[V_G(\mbf{p},\mbf{p}\,')\,\Phi(\mbf{p}\,')\right]
= {\cal V}_G((\mbf{p}-\mbf{p}\,')^2)} \nonumber \\ && \cdot \left[
\gamma^0 \Phi(\mbf{p}\,') \,\gamma^0
-\frac{1}{2}\,\left(
\mbf{\gamma} \Phi(\mbf{p}\,')\, \mbf{\gamma} +
(\mbf{\gamma}\hat{x}) \Phi(\mbf{p}\,')\, (\mbf{\gamma}\hat{x})\,
\right)\,\right]
\label{Couleq}
\end{eqnarray}
with the operator $\hat x={\bf x}/|{\bf x}|$, and
\begin{equation}
{\cal V}_G(\mbf{q}^{\,2}) =
4\pi\,\frac{4}{3}\,\frac{\alpha_s(\mbf{q}^{\,2})}{\mbf{q}^{\,2}}
\label{vgluon}
\end{equation}
For $\alpha_s(q^2)$ we make the following assumption. For
\(Q^2\equiv-q^2=\mbf{q}^2\gg\Lambda_{QCD}^2\)
the strong coupling \(\alpha_s(\mbf{q}^{\,2})\)
behaves like the running coupling constant
\(\alpha_s^{run}(Q^{\,2})\) of QCD, viz. \cite{pdg}
\begin{equation}
\alpha_s^{run}(Q^2) =
\frac{A}{\ln(Q^2/\Lambda_{QCD}^2)}
\,\left( 1 -  \,B\, \frac{\ln\,(\ln(Q^2/\Lambda_{QCD}^2))}
       {\ln(Q^2/\Lambda_{QCD}^2)} \right)
 + \; \ldots \label{alfas}
\end{equation}
with $A=4\pi/9$ and $B= 64/81$.
For \(Q^2 \ll \Lambda_{QCD}^2\) we assume some saturation value
\(\alpha_{sat}\) which will be a fit parameter of the model.
In between we simply assume some smooth interpolation between the two
 regions.

For consistency with the confining potential, we use co-ordinate
representation of the the Coulomb kernel. It may be parameterized as
follows \cite{jres94}.
\begin{equation}
  \alpha_s(r) = \frac{A} {2\,\ln \left(e^{-(\gamma+\mu a)}/a +
    e^{A/(2\alpha_{sat})} \right)} \,
  \left[1-B\,\frac{\ln\,(2\,\ln(e^{-\tilde{\mu}a}/a + e^{1/2}))}
    {2\,\ln(e^{-\mu a}/a + e^{B/2})} \right] \label{alfasr}
\end{equation}
where $\gamma = 0.577215\dots$ is the Euler Mascheroni constant, and
$a=r\Lambda_{QCD} $. We set \(\mu=4\) and
\(\tilde{\mu}=20\).

As discussed in \cite{jres94, mue94} the potential needs to be
regularized, which introduces an additional parameter $r_0$. The
reason for that is different from the non-relativistic case, where the
terms of order \(\mbf{p}^{\,2}/m^2\) like the spin-spin and spin-orbit
interaction lead to a collapse of the wave-function at the origin.  For
most of the Salpeter amplitudes and a fixed coupling constant Murota
\cite{mur83} has explicitly shown that the amplitudes are divergent for
\(r \rw 0\).  For a running coupling constant this divergence
is less pronounced, but still present. We therefore will use the
following regularized potential (in co-ordinate space)
\begin{eqnarray}
{\cal V}_G^F(r) &=& -\frac{4}{3}\,\frac{\alpha_s(r)}{r}
\;\;\;\;\mbox{for}\;\;r > r_0 \nonumber \\
{\cal V}_G^F(r) &=& a_G\,r^2+b_G
\;\;\;\;\mbox{for}\;\;r \le r_0 \label{VGFr}
\end{eqnarray}
where \(a_G\) and \(b_G\) are simply
determined by the condition that \({\cal
  V}_G^F(r)\) and its first derivative are continuous functions.  The
dependence of \(\alpha_s(r)\) on \(\Lambda_{QCD}\)  given
by eq.(\ref{alfasr}) is rather weak and may be compensated for by
modifying \(\mu\) and \(\alpha_{sat}\). We use
\(\Lambda_{QCD}=200\,MeV\) for our calculation.
The dependence of the mass spectra on the regularization parameter
\(r_0\) is very weak so that the differences in the mass spectra
calculated with the regularized and unregularized potential are quite
small. For our further calculation we will take \(r_0=0.1 \,fm\).

To solve the Salpeter equation numerically, eq. (\ref{salpeter}) is
rewritten as an eigenvalue problem, see e.g. \cite{jres94, lag92}.
This way it is it possible to utilize the variational principle to
find the respective bound states. To this end the Salpeter amplitude
$\Phi$ is expanded into a reasonable large number of basis states used
as a test function. As a suitable choice of basis states we have taken
Laguerre polynomials and found that about ten basis states lead to
sufficient accuracy, see also~\cite{jres94, mue94}.

\begin{figure}
\centering
  \leavevmode
  \epsfxsize=1.0\textwidth
  \epsfysize=15cm
  \epsffile{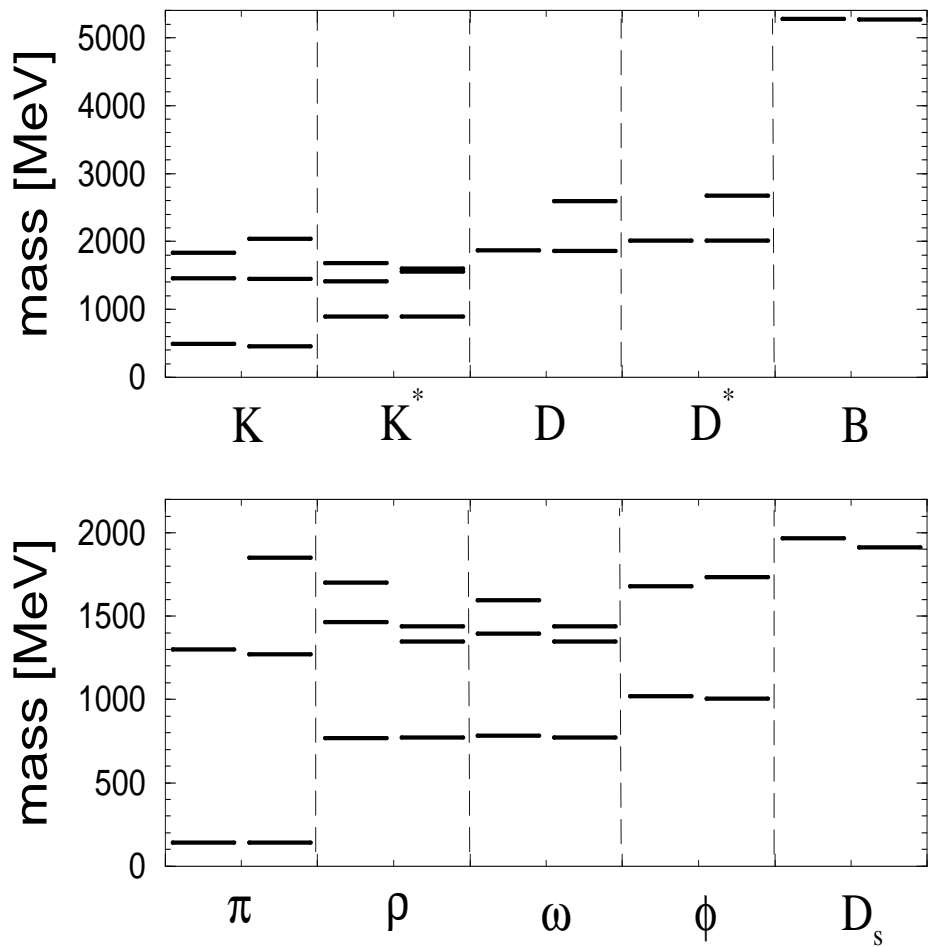}
\caption{
Meson mass spectrum for low radial excitation. Left column
experimental, right column model results, resp.}
\label{spectra}
\end{figure}

The parameters of the model are given in Table \ref{parameters}. These
are the quark masses, the offset \(a_c\) and slope \(b_c\) of the
confinement interaction eq. (\ref{vconf}) and the saturation value
\(\alpha_{sat}\) for \(\alpha_s(r)\) in eq.(\ref{alfasr}).  They are
determined to give a good overall description of the meson mass
spectrum. The mesons relevant for semileptonic decays, and their low
radiative excitations are shown in Figure \ref{spectra}.

\section{Form factors and semileptonic decays}

Semileptonic decays are treated in current-current approximation.  For
a transition $b\rw c$ the Lagrangian is given by
\beq
\label{lagrangian}
\CL_{cb}=\frac{G_F}{\sqrt{2}}\;V_{cb}\;\;h_{cb}^\mu\;j_\mu
\eeq
with the Cabbibo-Kobayashi-Maskawa matrix element $V_{cb}$.
The leptonic  $j_\mu$ and hadronic currents $h_{cb}^\mu$ are
defined by
\bqn
j_\mu&= &{\ovl \ell} \gamma_\mu (1-\gamma_5) \nu_\ell\\
h_{cb}^\mu&=&{\ovl c} \gamma^\mu (1-\gamma_5) b
\label{curhad}
\eqn
The relevant transition amplitudes $\bramket {D^{(*)}} {h_{cb}^\mu(0)}
{B}$ for $B\rw D$ and $B\rw D^*$ of the hadronic current can be
decomposed due to the Lorentz covariance of the current, thus
introducing form factors.  We use a standard representation in terms
of $F_0(q^2)$, $F_1(q^2)$ for $0^- \rw 0^-$ transitions, and $V(q^2)$,
$A_0(q^2)$, $A_1(q^2)$, $A_2(q^2)$ for $0^- \rw 1^-$ transitions.  The
exact definitions, and further references have been given
e.g. in~\cite{res94}. Note that $m_\ell^2\leq q^2\leq
q_{max}^2=(m_B-m_{D^{(*)}})^2$ due to kinematical reasons. Helicity
amplitudes $H_\pm$ and $H_0$ in terms of the  form factors
have been given by K\"orner and Schuler (KS) in a series of
papers~\cite{koe88} and are compiled by the particle data
group~\cite{pdg}. The respective decay rates into specific
helicity states $\Gamma_\pm$, $\Gamma_0$
are also given in the literature, see e.g.~\cite{pdg}.

\begin{figure}
\centering
  \leavevmode
  \epsfxsize=0.8\textwidth
  \epsffile{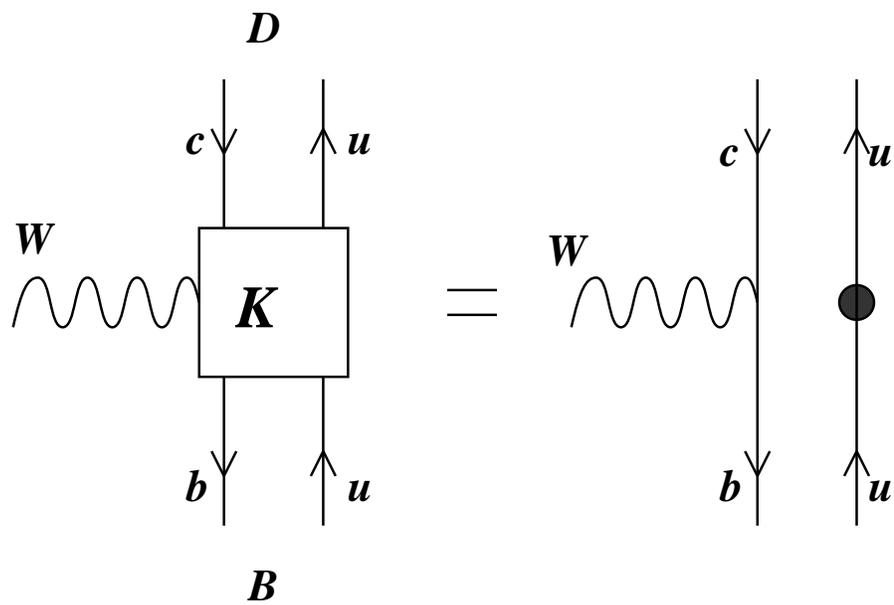}
\caption{\label{kernel}
Pictorial demonstration of one particle approximation of the
irreducible interaction kernel eq.(\protect{\ref{weakkernel}}), filled
circle denotes the inverse quark propagator }
\end{figure}

To determine the form factors from the model, we follow the general
prescription by Mandelstam \cite{man55}, see e.g. \cite{lur68} for a
textbook treatment.  The lowest order weak kernel reads (consider
e.g. the anti-quark current, flavour indices suppressed)
\begin{equation}
\label{weakkernel}
  K_{weak}^{\mu}(P,q,p,p') = \gamma^\mu (1-\gamma_5)
\,\left[{S^F_q}(P/2+p)\right]^{-1}\,\delta(p'-p-q/2)
\end{equation}
where \(p\) and \(p'\) denote the relative momenta of the incoming and
outgoing \(q\bar{q}\) pair, \(q=P-P'\) is the momentum transfer. A
pictorial demonstration of this approximation is given in
Figure~\ref{kernel}. The quark Feynman propagator is denoted by
$S^F_q$. The Dirac coupling to point-like particles is consistent with
the use of free quark propagators. Thus the semileptonic current
matrix element, e.g for $B\rw D$ that needs to be
calculated, reads
\begin{eqnarray}
\label{transitioncurrent}
  \lefteqn{\left\langle\,D, P_{D}\,\left|\,
h^\mu_{cb}(0)\,
  \right|\,B, P_B\,\right\rangle =} \\ &= & -
\;\int\!\!\frac{d^4p}{(2\pi)^4}\;\; tr \; \left\{
\bar{\Gamma}_{P_D}(p-q/2)\;{S^F_{{\ovl q}'}}(P_B/2+p-q)\;
\gamma^{\mu}(1-\gamma_5)\;\right.\nonumber\\
&&\left. \qquad \qquad\qquad{S^F_{\ovl q}}(P_B/2+p)\;
\Gamma_{P_B}(p)\;{S^F_q}(-P_B/2+p)\right\}
\nonumber
\end{eqnarray}
where $\Gamma_P(p)$ is the
amputated BS amplitude or vertex function
\beq
 \Gamma_P(p)  :=
[S^F_{q}(p_{q})]^{-1} \,\chi_P(p)\;[S^F_{\ovl q}(-p_{\ovl q})]^{-1}
\eeq
It may be computed in the rest frame from the equal time amplitude
$\Phi(\mbf{p})$ using the  Bethe-Salpeter equation
\begin{eqnarray}
  \Gamma_P(p)\left|_{_{P=(M,\mbf{0}\,)}}\right.  \; =\;
\Gamma(\mbf{p}\,) \; =\;
  -i\! \int\!\! \frac{d^3p'}{(2\pi)^4}
  \left[ V(\mbf{p},\mbf{p}\,')\Phi(\mbf{p}\,')\right]
\label{vert}
\end{eqnarray}
Finally, using Lorentz transformation properties,
 we can calculate the full BS
amplitude in any reference frame as
\begin{equation}
  \chi_P(p) = \;
  S_{\Lambda_P}^{}\;\;\chi_{(M,\mbf{0})}(\Lambda_P^{-1}p)\;\;
  S_{\Lambda_P}^{-1}.
\label{boo}
\end{equation}
where   $\Lambda_P$ is the pure Lorentz boost, and $S_{\Lambda_P}$ the
corresponding transformation matrix for Dirac spinors.

Due to the reconstruction of the full Bethe-Salpeter amplitude sketched
above, the transition matrix element eq.~(\ref{transitioncurrent}) is
manifestly covariant.

The spin part of the current is evaluated by a standard trace
technique appropriate for the particle anti-particle formalism.  The
radial part have been expanded in a basis of eleven Laguerre
functions.  The results are found to be stable within a large range of
the scale parameter of the basis. The matrix elements are then
compared with the parameterizations given above to determine the
semileptonic transition form factors.

\section{Semileptonic $B$ decays}

\begin{figure}
\centering
  \leavevmode
  \epsfxsize=0.8\textwidth
  \epsffile{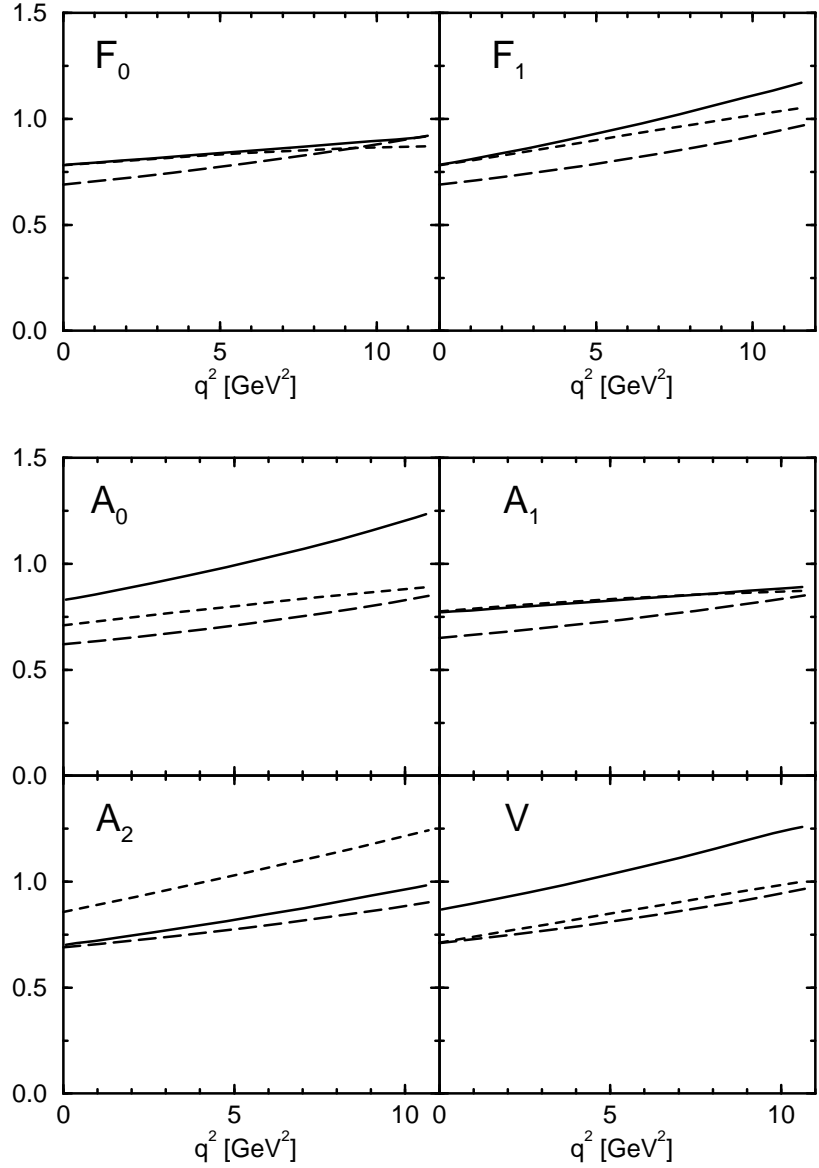}
\caption{A comparison of form factors for B $\rw$ D and B
  $\rw D^*$ transitions; our result (solid line), non
  relativistic result with relativistic corrections (dashed), BSW
(long-dashed)}
\label{BDformfactor}
\end{figure}

In Fig.~\ref{BDformfactor} we give the form factors relevant for the
transition of $B\rw D$ and $B\rw D^*$. They are calculated using the
Salpeter amplitudes reproducing the meson mass spectra as shown in
Fig.~\ref{spectra}. Our calculation is given by the solid line. The
mono-pole dominance ansatz of Bauer, Stech and Wirbel (BSW)
\cite{bjo89} is shown as a long dashed line.  The short dashed line
shows an earlier result within the non-relativistic framework
(CQM). The latter, however, includes relativistic corrections in the
current operators~\cite{res94}.

In Table~\ref{formmax} we explicitly give the ratios of the form
factors $A_2/A_1$ and $V/A_1$ at $q^2=q^2_{max}$. The values fit
experimental data, which however, have rather large error bars. For
comparison we have also given other model values as compiled
in~\cite{pdg}, in particular of the improved non-relativistic
constituent quark model given earlier~\cite{res94}.

\begin{figure}
\centering
  \leavevmode
  \epsfxsize=1.0\textwidth
  \epsfysize=20cm
  \epsffile{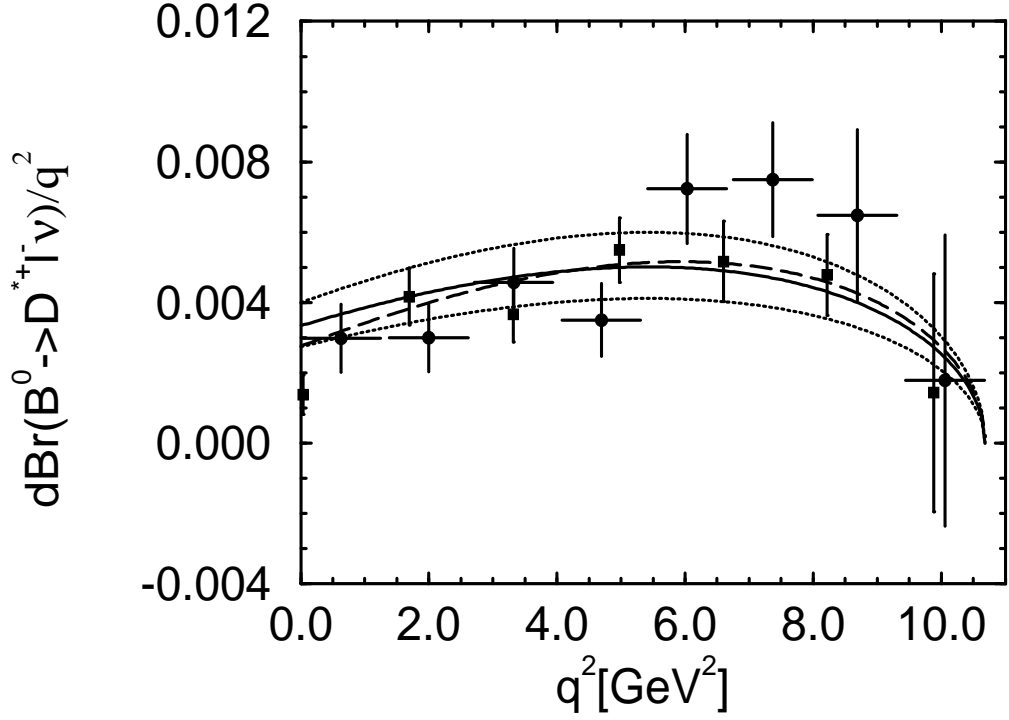}
\caption{
  $q^2$ distribution of $B \rw D^{*+}\ell^-{\ovl \nu}$.
  Experiments given by ARGUS (circles) and CLEO (squares).  The solid
  line calculated with $V_{cb}=0.032$ and life time $\tau_B=1.49ps$;
  the upper and lower dotted lines with $V_{cb}=0.032\pm 0.003$
  respectively. The dashed line corresponds to the nonrelativistic
  treatment with  $V_{cb}=0.034\pm 0.003$. }
\label{decay}
\end{figure}

The exclusive decay spectrum is shown in Fig.~\ref{decay}. We
compare our results to recent ARGUS~\cite{alb93} and CLEO data
\cite{cas93,san93}. We find a best fit with $V_{cb}=(0.032\pm 0.003)
(\tau_B/1.49ps)^{1/2}$.  The error resulting from $\chi^2$ fitting is
indicated by the upper and lower dotted line. Our previous result
using a non-relativistic formalism with relativistic corrections in the
current operators is given by the dashed line. We find
$V_{cb}$ consistent with each other \cite{res94}.

The resulting total branching ratios for semileptonic $B$ decays agree
well with experimental data given by the Particle Data Group
\cite{pdg}, and by the CLEO collaboration~\cite{san93}
and ARGUS collaboration~\cite{alb93}. They are given in
Table~\ref{Bbranching}.

In Table~\ref{afb} we give the resulting forward backward asymmetry
$A_{FB}$ and the asymmetry parameter $\alpha$ defined
e.g. in~\cite{koe88} respecting the lepton cut-off momentum of the
CLEO collaboration.  The forward backward asymmetry sensitive to
parity violation is defined through
\beq
A_{FB} = \frac{N_F-N_B}{N_F+N_B}
\eeq
with $N$ the number of leptons in forward, resp. backward hemisphere in
the rest system of the $W$-boson. For $A_{FB}$ we have used a
symmetric cut on the lepton momentum $p_\ell$ as utilized by the CLEO
collaboration~\cite{san93} for technical reasons.

The helicity alignment $\alpha$ describes the $D^{*+}$ polarization
extracted from the $D^{*+}\rw D^0\pi^+$ decay angle
distribution $W(\theta^*)$, viz.
\beq
W(\theta^*)\propto 1+\alpha \cos^2\theta^*
\eeq
For $\alpha$ only a lower cut has been introduced. Both observables
are in good agreement with experimental results.

\begin{figure}
\centering
  \leavevmode
  \epsfxsize=\textwidth
  \epsffile{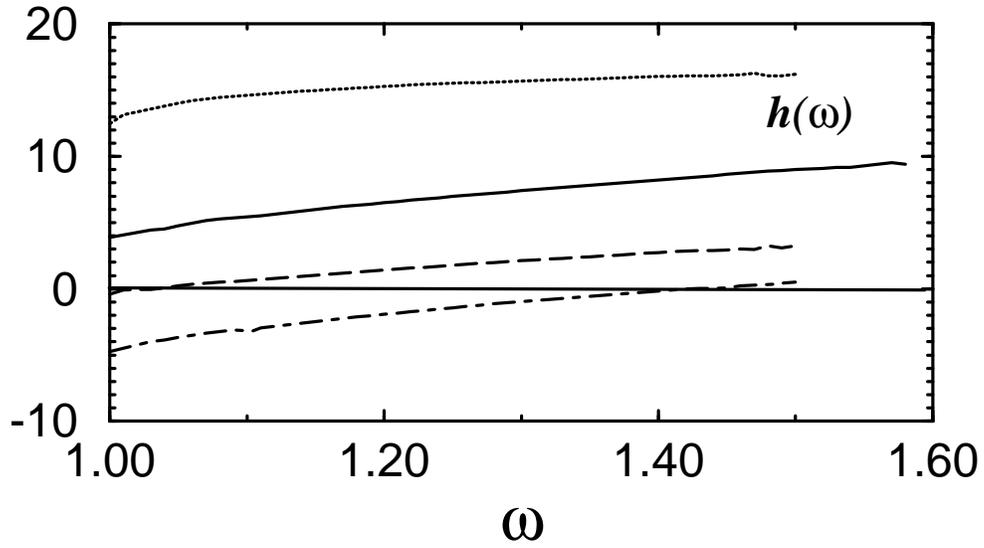}
\vspace*{-3cm}
\caption{\label{isgw}
Deviation of the model form factors from the Isgur Wise
function in [\%].  From top to bottom: $h_V$, $h_+$,
  $h_{A_1}$, $h_{A_3}$. }
\end{figure}

We would now like to connect our results to the notion of heavy quark
symmetry. In this context we reparametrize the form factors in terms
of $h_{+}(\omega)$, $h_{-}(\omega)$ for $0^-\rw 0^-$, and
$h_V(\omega)$, $h_{A_{1}}(\omega)$, $h_{A_{2}}(\omega)$,
$h_{A_{3}}(\omega)$ for $0^-\rw 1^-$ transitions, with
\bqn
\omega & = & \frac{m_B^2 + m_{D^{(*)}}^2 - q^2}{2 m_B m_{D^{(*)}}}
\eqn
The transformation equations are given in~\cite{res94}. The advantage
of this parameterization is that in the heavy quark mass limit
($m_{c,b} \rw\infty$)
\bqn
h_V(\omega)=h_{A_1}(\omega)=h_{A_3}(\omega)=h_+(\omega)
&=&\xi(\omega)\label{ffisgw}\\
h_{A_2}(\omega)=h_-(\omega)&=&0
\eqn
where $\xi(\omega)$ is a universal function known as Isgur Wise
function~\cite{isg90}.  The heavy quark mass limit has been performed
numerically by multiplying $m_{c,b}$ with a large factor, while
keeping all other parameters as given in Table~\ref{parameters}. The
meson amplitudes to evaluate the transition matrix
elements~(\ref{transitioncurrent}) are then calculated by
diagonalizing the equivalent eigenvalue problem. Due to numerical
reasons the heavy quark masses cannot be chosen too large.  However, a
suitable choice leads to converging form factors of
eq. (\ref{ffisgw}), which differ by less than 0.1\%. This resulting
function is then defined to be the Isgur Wise function
$\xi_{IBS}(\omega)$. Its values are shown in Table~\ref{isgw}.

We find that is it possible to approximately parameterize this function
by either a pole or an exponential fit, viz.
\bqn
\xi_{pole}(\omega)&\simeq
&\frac{\Lambda^2+1}{\Lambda^2+\omega}\\
\xi_{exp}(\omega)&\simeq &\exp\left[\alpha^2 (1-\omega)\right]
\eqn
Choosing $\Lambda=0.14$, or $\alpha^2=0.80$ the deviation of this fits
from the calculated values $\xi_{IBS}(\omega)$ are smaller or about
1\%, whereby $\xi_{exp}(\omega)$ leads to a slightly better overall fit.

In Figure~\ref{isgw} we show the deviations of the IBS form factors,
$h_V(\omega)$, $h_{A_1}(\omega)$, $h_{A_3}(\omega)$, $h_+(\omega)$,
(viz. without implementing the `numerical' heavy quark mass limit)
from the ideal heavy quark limit form factor $\xi(\omega)$. The form
factor $h_V(\omega)$ deviates rather strongly from the Isgur Wise
function, whereas $h_+$ and $h_{A_3}$ are within 10\%. Although our
result is not a $\Lambda_{QCD}/m_q$ expansion, it is interesting to
note, that $h_{A_1}(1)=0.995$ agrees with the expectation of Luke's
theorem~\cite{luk90}, and $h_{A_1}(\omega)$ is rather close to the
limiting case $\xi_{IBS}(\omega)$.

\section{Semileptonic $D$ decays}

\begin{figure}
\centering
  \leavevmode
  \epsfxsize=0.8\textwidth
  \epsffile{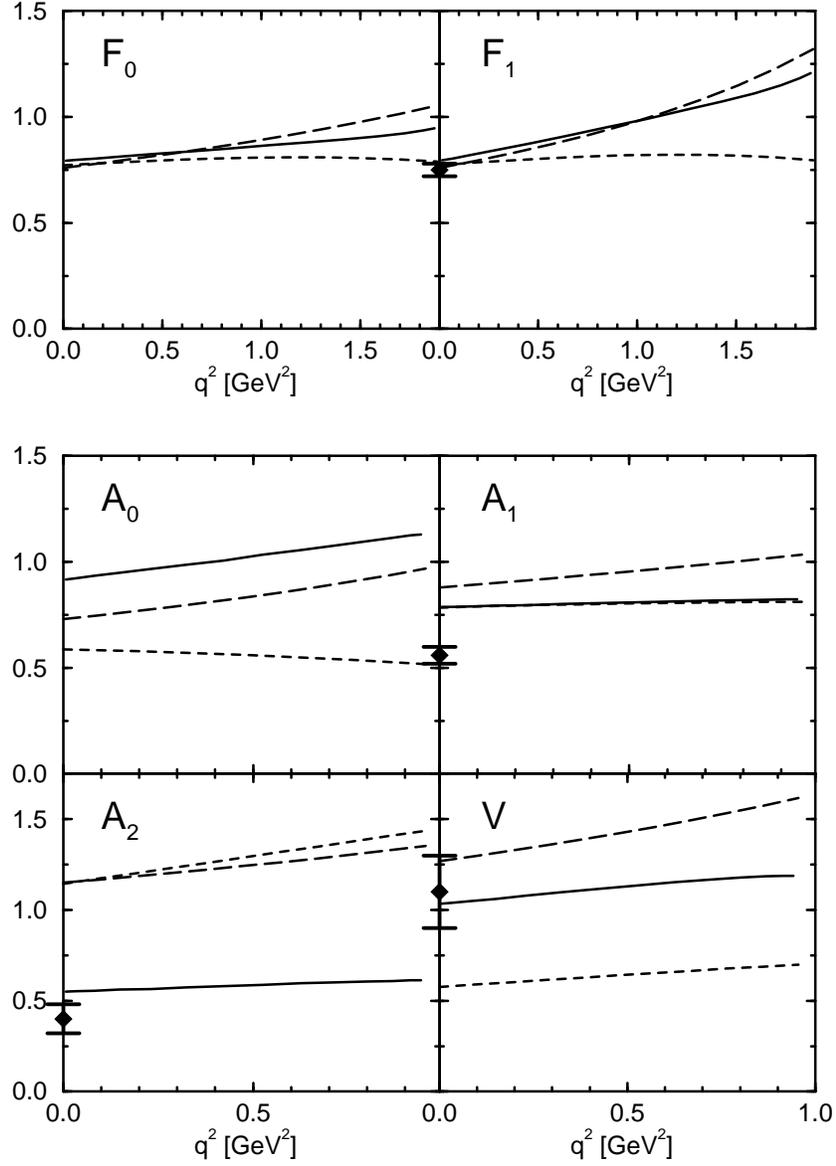}
\caption{\label{DKformfactor}
  A comparison of form factors for D $\rw$ K and D
  $\rw $ $K^*$ transitions; our result (solid line), non
  relativistic result with relativistic corrections (dashed),
  BSW (long--dashed),
  empirical form factors (shaded area, where given)}
\end{figure}

In Figure \ref{DKformfactor} we show the $D\rw K^{(*)}$ form
factors. The results of the IBS model are compared to other models as
well as to empirical data. The empirical data points at $q^2=0$ are
given by the particle data group assuming a pole dominance behaviour of
the form factors~\cite{pdg}.  Note that the covariant treatment (solid
line) can reproduce a pole-like behavior in contrast to the
non-relativistic calculations (dashed line).

Except for the form factor $A_1(0)$ the model form factors agree
reasonably well with the empirical values given. It is remarkable that
$A_2(0)$ turns out to be much closer to the empirical value than the
non-relativistic result or the BSW values that are fixed by quark wave
functions in the infinite momentum frame. On the other hand $A_1(0)$
is too large by 50\%.

In Table~\ref{Dbranching} we show the resulting partial decay rates of
$D\rw K^{(*)}$, $D_s\rw\Phi$. Although the decay into $K$ reproduces
the experimental value nicely, the decays into $K^*$ and $\Phi$ are
rather off the experimental data. In addition  the ratio of helicity
rates $\Gamma_+/\Gamma_-$ shown in Table~\ref{Gamma} turns out larger
than the experimental ones. This reflects the difference in $A_1(0)$
between the experimental and empirical values.  Table~\ref{Gamma} also
shows the ratio of longitudinal to transverse polarized rates
($\Gamma_L/\Gamma_T=\Gamma_0/(\Gamma_-+\Gamma_+)$), which is well
reproduced by the model.

The decay into $\eta$ is not calculated since this requires a
discussion of $\eta,\eta'$ mixing which is not touched here.  At the present
stage it would require additional assumptions and the introduction of
mixing parameters, which we like to exclude here.  Such mixing may be
generated in a natural way though instanton effects
\cite{tho76}, or two gluon exchanges. The consequences of instanton
effects have been studied in~\cite{jres94, mue94}.

\section{Non-leptonic decays}

Due to strong interaction non-leptonic weak decays provide additional
phenomena. Examples are hard gluon exchanges, quark rearrangement,
annihilation and long range effects. Thus, extraction of fundamental
physical constants such as the Cabbibo-Kobayashi-Maskawa matrix
elements is more difficult than in the semileptonic case.  Discussion
on QCD corrections have been triggered by the recent CLEO data on $B$
meson decays. The r\^ole of QCD correction needs clarification,
however, such a discussion goes beyond the scope of the present paper.  For
definiteness we shall neglect all gluonic effects in the weak
operators. Implicitly some gluonic effects are included in the Bethe
Salpeter amplitudes via the interaction kernel. Therefore we also need
not to worry about double counting etc. The simple weak interaction
kernel for non-leptonic decay may then be written in current
approximation as
\bqn
\label{efflagrangian}
\CL_{eff} & = & -\frac{G_F}{\sqrt{2}} \sum_{\alpha^\prime\beta^\prime
\alpha\beta}
 V_{\alpha^\prime \beta^\prime} V^*_{\alpha\beta}
h_{\mu,\alpha\beta}h^\mu_{\alpha^\prime \beta^\prime }
\eqn
with $\alpha,\alpha^\prime \in\{u,c\}$, $\beta,\beta^\prime
\in\{d,s\}$. The Lagrangian (\ref{efflagrangian}) is then evaluated between the
meson amplitudes in the same way as for semileptonic decays, given in
(\ref{transitioncurrent}).

\begin{figure}
\centering
  \leavevmode
  \epsfxsize=0.8\textwidth
  \epsffile{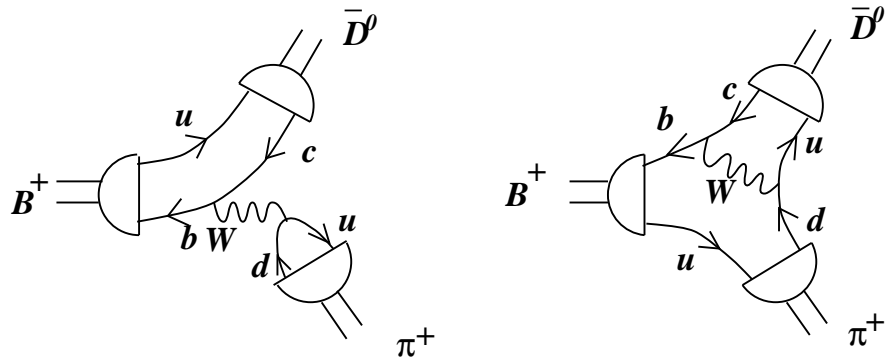}
  \vspace*{2cm}
\caption{\label{nonlepfig}
Type I (left) and Type II (right) nonleptonic decays. In the
limit of heavy W-Boson, Type II decays can be Fierz transformed. }
\end{figure}

Note that due to Fierz rearrangement two generic types of
contributions are possible. These are demonstrated in
Figure~\ref{nonlepfig}.  The generic form of class I is given by the
following eq. (\ref{hypfactor}) those of class II by
eq. (\ref{ahypfactor}), class III are mixed forms of both.
\bqn
\bramket{\pi^+\;D^-}{h_{\mu,ud} h^\mu_{cb}} {B^0}
&\rw
&\bramket{\pi^+}{h_{\mu,ud}}{0} \bramket{D^-}{h^\mu_{cb}}{B^0}
\label{hypfactor}\\
\bramket{\pi^0\;D^0}{h_{\mu,ud}h^\mu_{cb}}{B^0}
&\rw
&\frac{1}{3}\bramket{D^0}{h_{\mu,cd}}{0}\bramket{\pi^0}{h^\mu_{ub}}{B^0}
\label{ahypfactor}
\eqn
The factor $1/3$ stems from the  colour suppression of the Fierz
rearranged diagram.

In order to separate the physical effects, we have used empirical
values for the vacuum transition part $\bramket{X}{h_\mu}{0}$ to
evaluate the non-leptonic transition matrix elements. The empirical
values for the decay constants used in the calculation of non-leptonic
decays along with the calculated ones from the IBS model are shown in
Table~\ref{decayconstants}. They are not so well reproduced by the IBS
model, however calculated values are by orders of magnitudes closer to
the empirical ones compared to the non-relativistic approach.

Experimental values of decay constants are available only for $\pi$
and $K$ from leptonic weak decays. For the $D$-meson decay constant
only an upper limit exists, $f_{D^+}<310$MeV.  For $f_{D_s}$ we have
chosen $f_{D_s}=300$MeV, which is close to values found by
Rosner~\cite{ros90} and an ARGUS analysis using different types of
model analyses~\cite{alb92}. For $f_{\rho}$ we have used
$f_\rho=205$MeV, which has been suggested by~\cite{neu92}. Other decay
constants are taken from~\cite{bau87}.

We consider only those non-leptonic decays where experimental values
are known. These are taken from the compilation of the Particle Data
Group~\cite{pdg}, which include also some (preliminary) results from
recent CLEO~\cite{san93} and ARGUS~\cite{alb92, alb93} collaborations.

Results are shown in Tables~\ref{B0nonlep} to \ref{Dsnonlep}. The
Cabbibo Kobayashi Maskawa matrix elements used and $V_{cb}=0.032$ and
others are given in \cite{pdg}. The description of $B$ meson
non-leptonic decays is rather satisfactory. The relative sign between the
amplitudes of type III decays consistent with the Lagrangian given in
eq. (\ref{efflagrangian}) is positive. This holds also for $D$ meson
decays. On the other hand, experimental values for type III $D$ decays
are not reproduced. This is usually attributed to gluonic
effects. However, we shall not elaborate on this point any further.

\section{Summary and Conclusion}

Utilizing the Bethe Salpeter equation we have calculated the meson
mass spectrum, as well as weak semileptonic decay observables of $B$
and $D$ mesons. The interaction kernel has been assumed instantaneous.
This way the Bethe Salpeter equation reduces to a (full) Salpeter
equation as given in eq.(\ref{salpeter}).  The interaction consists of
a one gluon exchange evaluated in the Coulomb gauge and a linear
confinement, which are given in co-ordinate space. The parameters of
the model are fixed to reproduce the meson mass spectrum.

To evaluate current matrix elements, and in order to make explicit
covariance transparent, the Bethe Salpeter amplitude
$\chi_P(p^0,\mbf{p})$ has been fully reconstructed. In this way it is
possible to extract form factors by comparison of the Lorentz
structure of the standard notation of the current.

To calculate the Isgur Wise function the heavy quark mass limit has
been achieved in a numerical sense. Within the model uncertainties the
resulting function may be parameterized by an exponential as well as by
a pole-like function. Deviation of the form factor $h_{A_1}(0)$ from
unity is very small as expected from Luke's theorem. Although it is
possible to extract the value of the Cabbibo Kobayashi Maskawa matrix
element $V_{cb}$ without knowing the exact shape of the Isgur Wise
function, its notion is desirable since it leads to more insight into
the dynamical mechanism of the $q\ovl q$ system.

{}From the semileptonic decay rate $B\rw D^* e\nu$ the Cabbibo Kobayashi
Maskawa matrix element turns out to be $V_{cb}=0.032\pm 0.003$, which
is close to the non-relativistic value. We have taken
$\tau_B=1.49ps$~\cite{pdg} for the $B$ meson life time.

As a main outcome of the covariant treatment we find, that the
empirical pole behavior of the form factors of the semileptonic $D$
decays is well reproduced. We consider this as an important advantage
with respect to the non-relativistic treatment, where this is not
possible.  Also empirical values are reasonably well reproduced except
for $A_1(q^2)$. The total decay rate for $D\rw Ke\nu$ is reproduced
rather well, whereas for $D\rw K^*e\nu$ and $D_s\rw \phi e\nu$
experimental values are overestimated. Since presently no model
appears able to explain all form factors, this may well have a more
profound reason than a simple shortcoming of the model. Changing the
model parameters given in Table~\ref{parameters} may lead to some
improvement on $A_1(q^2)$ however, the mesonic mass spectrum Figure
\ref{spectra} cannot be reproduced then.

As a further application we have calculated the non-leptonic decays
leaving the discussion of gluonic corrections and final state
interaction aside. We find reasonable results, however without solving
the relative sign problem of type III decays. This needs further
consideration, which is beyond our present intention.

In conclusion, we presented a covariant model to treat mesons in terms
of the underlying quark fields, respecting simple QCD phenomena such
as confinement and spin splitting. It allows to extract fundamental
quantities of the standard model from hadronic degrees of freedom in a
consistent way using Mandelstam formalism. Since the model leads to
reasonable results one may now also address more exotic phenomena, such
as penguin diagrams or rare decays.

\section{Acknowledgment}
One of us (M.B.) is grateful to Yura Kalinovsky for reading the
manuscript and a valuable discussion on the Isgur Wise function. This
work was supported in part by the Bundesminister f\"ur Forschung und
Technologie.

\clearpage


\begin{table}
\caption{Parameter of the model kernel }
\label{parameters}
\[
\begin{array}{cccccccc}
\hline\hline
m_{u,d} &m_s   &m_c     &m_b     &a_c     &b_c       &r_0
&\alpha_{sat}\\
\mbox{[GeV]}  &\mbox{[GeV]}  &\mbox{[GeV]}  &\mbox{[GeV]}
&\mbox{[GeV]} &\mbox{[GeV/fm]}& \mbox{[fm]} &\\[1ex]
\hline
\vspace{1ex}
0.200 &0.440 &1.738 &5.110 &-1.027 &1.700 &0.1  &0.391\\
\hline\hline
\end{array}
\]
\end{table}

\begin{table}[b]
\caption{Comparison of ratios of form factors for
  $B\rw {\ovl D}^*\ell^+\nu_\ell$ at $q^2=q^2_{max}$}
\label{formmax}
\[
\begin{array}{lcc}
\hline\hline\\
&(A_2/A_1)(q^2_{max})&(V/A_1)(q^2_{max})\\[1ex]
\hline
\vspace{1ex}
\mbox{CLEOII fit (a)} & 1.02\pm 0.24 & 1.07\pm 0.57\\
\mbox{CLEOII fit (b)} & 0.79\pm 0.28 & 1.32\pm 0.62\\
\mbox{IBS}            & 1.10         & 1.41\\
\mbox{CQM}            & 1.42         & 1.15\\
\mbox{BSW}            & 1.06         & 1.14\\
\mbox{KS}             & 1.39         & 1.54\\
\hline\hline
\end{array}
\]
\end{table}

\begin{table}[b]
\caption{Branching ratios for B decays. We use $\tau_B =1.49$ps, a)
recent CLEO, b) recent ARGUS data}
\label{Bbranching}
\[
\begin{array}{r@{}c@{}lcc}
\hline\hline\\
\multicolumn{3}{c}{\makebox{decay}}&Br_{IBS} [\%]&Br_{exp} [\%]\\[1ex]
\hline
\vspace{1ex}
{\ovl B}^0 &\rw  &D^{*+}\ell^-{\ovl \nu}_\ell &4.2 &4.9\pm 0.8\\
&&&&4.50\pm 0.44 \pm 0.44^a\\
&&&&5.2\pm 0.5\pm 0.6^b\\
{\ovl B}^0 &\rw  &D^{+}\ell^-{\ovl \nu}_\ell  &1.6 &1.6 \pm 0.7\\
\hline\hline
\end{array}
\]
\end{table}

\begin{table}[b]
\caption{Forward backward asymmetry and asymmetry parameter $\alpha$
  for the decay ${\ovl B}^0 \rw D^{*+}\ell^-{\ovl \nu}_\ell$.
CLEO with lepton cut-off $p^{cut}_\ell=1.0/1.4 \mbox{GeV}$, ARGUS
without cut-off}
\label{afb}
\[
\begin{array}{cccc}
\hline\hline\\
&p_\ell^{cut}&\makebox{IBS} &\makebox{experiment}\\[1ex]
\hline\\
A_{FB} &0  &0.20 &0.2 \pm 0.1 \\
       &1.0&0.14 &0.14\pm 0.07\\
\alpha &0  &1.37 &1.1 \pm0.4\\
       &1.4&0.50 &0.65\pm 0.7\\
\hline\hline
\end{array}
\]
\end{table}

\begin{table}[b]
\caption{Isgur Wise function calculated from the instantaeous Bethe
Salpeter equation as explained in the text}
\label{fit}
\[
\begin{array}{cc}
\hline\hline
\omega&\xi(\omega)\\
\hline
 1.00& 1.000\\
 1.05& 0.958\\
 1.10& 0.918\\
 1.15& 0.880\\
 1.20& 0.845\\
 1.25& 0.811\\
 1.30& 0.780\\
 1.35& 0.750\\
 1.40& 0.721\\
 1.45& 0.695\\
 1.50& 0.669\\
 1.55& 0.645\\
 1.60& 0.623\\
\hline\hline
\end{array}
\]
\end{table}

\begin{table}[b]
\caption{$D\rw K^{(*)}$ and $D_s\rw
  \phi$ decays}
\label{Dbranching}
\[
\begin{array}{r@{}c@{}lcc}
\hline\hline\\
\multicolumn{3}{c}{\makebox{decay mode}}
&\Gamma_{\mbox{IBS}} [10^{10}s^{-1}]
&\Gamma_{\mbox{exp}} [10^{10}s^{-1}]\\[1ex]
\hline\\
D &\rw  &{\ovl K}  e \nu_e       &8.8 &8.2 \pm 0.4\\
D &\rw  &{\ovl K^*}e \nu_e       &8.2 &4.6 \pm 0.4\\
\hline
D_s &\rw&\phi\ell^+{\ovl\nu}_\ell&7.7 &3.11 \pm 1.11\\
\hline\hline
\end{array}
\]
\end{table}

\begin{table}[b]
\caption{$D\rw K^{(*)}$ helicity ratios}
\label{Gamma}
\[
\begin{array}{lcc}
\hline\hline\\
&\Gamma_L/\Gamma_T&\Gamma_+/\Gamma_-\\
\hline\\
\mbox{exp. average} & 1.23\pm 0.13&0.16\pm0.04\\
\mbox{IBS   }       & 1.38       &0.30       \\
\hline
\end{array}
\]
\end{table}

\begin{table} \label{decayconstants} \centering \caption{Empirical
decay constants $f_{emp}$ used for calculating nonleptonic decays
compared to calculated IBS values $f_{IBS}$}
\vspace{0.5cm}
\begin{tabular}{ccc}
\hline\hline\\
meson & $f_{emp}$ & $f_{IBS}$ \\
type & [MeV] & [MeV] \\
\hline \\
$\pi^+$ & 132 & 153\\
$\pi^0$ & 93 &108\\
$K^\pm$ & 162 & 220\\
$\bar K^0$ & 162 & 220\\
$D^+$ & 220 &293\\
$D^0$ & 220 & 293\\
$D_s^+$ & 300 & 342\\
$\rho^+$ & 205 & 480\\
$\rho^0$ & 145 & 339\\
$K^{*-}$ & 220 & 503\\
$\bar K^{*0}$ & 220 &503\\
$D^{*+} $ & 220 & 409\\
$D^{*0}$ & 220 & 409\\
$D_s^{*+}$ & 300& 467\\
$J/\Psi$ & 382 & 571\\
$a_1^+$ & 220 & 371\\
 \hline\hline
\end{tabular} \end{table}

\begin{table}[htbp]
\caption{\label{B0nonlep}
Decay widths of weak nonleptonic $B^0$ decays in units of $s^{-1}$.
Upper part type I, lower type II decays}
\vspace{0.5cm}
\centering
\begin{tabular}{lll}
\hline\hline\\
$B^0$ decays & $\Gamma_{IBS}$ & $\Gamma_{exp}$ \\[1ex]
 \hline\\
$D^- \pi^+$         & 0.208 & $0.20 \pm 0.03$ \\
$D^- \rho^+$        & 0.465 & $0.52 \pm 0.09$ \\
$D^- D_s^+$         & 0.840 & $0.53\pm 0.27$ \\
$D^{*-} D_s^+$      & 0.667 & $0.80 \pm0.40$ \\
$D^- D_s^{*+}$      & 0.451 & $1.40 \pm 1.0$ \\
$D^{*-} D_s^{*+}$   & 0.914 & $1.33 \pm 0.80$ \\
$D^{*-} \rho^+$      & 0.522 & $0.49 \pm 0.10$ \\
\hline
$K^0 J/\Psi$        & 0.033 & $0.05 \pm 0.01$ \\
$K^{*0}J/\Psi$      & 0.075 & $0.11 \pm0.02$ \\
\hline\hline
\end{tabular}
\end{table}

\begin{table}[htbp]
\caption{\label{B+nonlep}
Decay widths of weak nonleptonic $B^+$ decays in units of $s^{-1}$.
{}From upper to lower part: type I, II, III decays, resp.}
\vspace{0.5cm}
\centering
\begin{tabular}{lll}
\hline\hline\\
$B^+$ decays & $\Gamma_{IBS}$   & $\Gamma_{exp}$  \\[1ex]
 \hline\\
$\bar D^0 D_s^+$    & 0.840 & $1.10 \pm 0.39$ \\
$\bar D^0 D_s^{*+}$ & 0.451 & $0.78 \pm 0.65$ \\
$\bar D^{*0} D_s^+$ & 0.667 & $0.65 \pm 0.45$ \\
$\bar D^{*0} D_s^{*+}$ & 0.914 & $1.56 \pm 0.84$ \\
\hline
$K^+ J/\Psi$        & 0.033 & $0.07 \pm 0.01$ \\
$K^{*+} J/\Psi$     & 0.075 & $0.11 \pm 0.03$ \\
\hline
$\bar D^0 \pi^+$    & 0.217 & $0.34 \pm 0.03$ \\
$\bar D^0 \rho^+$   & 0.507 & $0.87\pm 0.12$\\
$\bar D^{0*} \pi^+$ & 0.307 & $0.34 \pm 0.05$\\
$\bar D^{0*} \rho^+$& 0.561 & $1.01\pm 0.20$ \\
\hline\hline
\end{tabular}
\end{table}

\begin{table}[htbp]
\caption{\label{D0nonlep}
Decay widths of weak nonleptonic $D^0$ decays  in units of
$s^{-1}$. Upper part type I, lower part type II decays.}
\vspace{0.5cm}
\centering
\begin{tabular}{lll}
\hline\hline\\
$D^0$ decays       & $\Gamma_{IBS}$ & $\Gamma_{exp}$  \\[1ex]
 \hline\\
$K^- \pi^+$         & 10.52 & $9.66 \pm 0.34$ \\
$K^- \rho^+$        & 15.79 & $25.06 \pm 3.13$ \\
$\bar K^0 \phi$     & 0.58  & $2.00 \pm 0.29$ \\
$K^- a_1^+$         & 2.21  & $19.04 \pm 2.89$ \\
$K^{*-} \pi^+$      & 7.87  & $11.81 \pm 1.45$ \\
$K^{*0} \pi^0$      & 0.002 & $7.23 \pm 0.96$ \\
$K^{*-} \rho^+$     & 11.29 & $14.22 \pm 5.78$ \\
$\pi^+ \pi^-$       & 0.14  & $0.38 \pm 0.03$ \\
$K^+ K^-$         & 0.78  & $1.09 \pm 0.07$ \\
$K^0 \bar K^0$    & 0.78  & $0.27 \pm 0.10$ \\
$K^{*+} K^-$      & 0.43  & $0.82 \pm 0.19$ \\
$K^{*-} K^+$      & 0.43  & $0.43 \pm 0.24$  \\
$K^{*0}\bar K^{*0}$& 0.40  & $0.70 \pm 0.39$ \\
\hline
$\bar K^0 \pi^0$    & 0.001 & $4.94 \pm 0.63$ \\
$\bar K^0 \rho^0$   & 3.00  & $2.65 \pm 0.43$ \\
$\pi^0 \pi^0$     & 0.07  & $0.21 \pm 0.06$ \\
$\phi \rho^0$     & 0.04  & $0.46 \pm 0.12$ \\
\hline\hline
\end{tabular}
\end{table}

\begin{table}[htbp]
\caption{\label{D+nonlep}
Decay widths of weak nonleptonic $D^+_{(s)}$ decays  in units of
$s^{-1}$. Upper part type I, lower type III decays}
\vspace{0.5cm}
\centering
\begin{tabular}{lll}
\hline\hline\\
$D^+$decay & $\Gamma_{IBS}$ & $\Gamma_{exp}$ \\[1ex]
 \hline\\
$\bar K^0 a_1^+$    & 2.21  & $7.66 \pm 1.61$ \\
$\bar K^0 K^+$      & 0.78  & $0.74 \pm 0.16$ \\
$\phi \pi^+$        & 0.031 & $0.63 \pm 0.08$ \\
$\bar K^{*0} K^+$   & 0.43  & $0.48 \pm 0.09$ \\
$K^{*+} \bar K^{*0}$& 0.40  & $2.46 \pm 1.04$\\
\hline
$\bar K^0 \pi^+$    & 11.25 & $2.59 \pm 0.27$ \\
$\bar K^0 \rho^+$   & 17.11 & $6.24 \pm 2.37$ \\
$\bar K^{*0} \pi^+$ & 7.91  & $2.08 \pm 0.38$ \\
$\pi^+ \pi^0$       & 2.07  & $0.24 \pm 0.07$ \\
\hline\hline
\end{tabular}
\end{table}

\begin{table}[htbp]
\caption{\label{Dsnonlep}
Decay widths of weak nonleptonic $D^+_{(s)}$ decays  in units of
$s^{-1}$. Upper part type I lower part type II decays}
\vspace{0.5cm}
\centering
\begin{tabular}{lll}
\hline\hline\\
$D_s^+$decay & $\Gamma_{IBS}$ & $\Gamma_{exp}$ \\[1ex]
 \hline\\
$K^+ \bar K^{*0}$ & 1.45  & $7.49 \pm 1.50$ \\
$\phi \pi^+$      & 7.35  & $7.49 \pm 0.86$ \\
$\phi \rho^+$     & 10.68 & $13.92 \pm 3.43$ \\
\hline
$K^+ \bar K^{*0}$ & 1.76  & $7.07 \pm 1.07$ \\
$K^{*+} \bar K^0$ & 0.93  & $8.99 \pm 2.14$\\
\hline\hline
\end{tabular}
\end{table}

\clearpage

\end{document}